\documentclass[12pt]{article}  
\usepackage{graphicx} 
\usepackage{wrapfig}  
\usepackage[utf8]{inputenc}
\usepackage[hidelinks]{hyperref}
\usepackage{xcolor}
\usepackage{amsthm}
\usepackage{amsmath,amssymb,amsfonts}
\usepackage[version=3]{mhchem}
\usepackage[utf8]{inputenc}
\usepackage[letterpaper, bindingoffset=0.0in, left=1in, right=1in, top=1in, bottom=1in]{geometry}
\usepackage{setspace} 
\usepackage{parskip} 
\usepackage{authblk}
\usepackage{blindtext}

\begin{document}

\title{\textbf{Statistical mechanical theory of liquid water}}
\author[1]{Lakshmanji Verma \thanks{Contributing author: lakshmanji.verma@stonybrook.edu}}
\author[1,2]{Ken A Dill \thanks{Corresponding author: dill@laufercenter.org}}

\affil[1]{\textit{{Laufer Center for Physical and Quantitative Biology}, {Stony Brook University}, {NY}, {USA}}}
\affil[2]{\textit{{Department of Physics $\And$ Department of Chemistry}, {Stony Brook University}, {NY}, {USA}}}


\maketitle

\begin{abstract}
Water is an unusual liquid. Its thermophysical properties are non-monotonic with temperature T and pressure p. It's not been known how water's behaviors are encoded in its molecules. We give a statistical physics model, \emph{Cage Water}, which assumes three bonding states: van der Waals, pairwise hydrogen bonding, and multi-body cooperative caging hydrogen bonds. The model is analytical, so very fast to compute, yet it gives excellent agreement with extensive pT experiments.  Through readily interpretable substates, Cage Water explains water's liquid anomalies -- including its controversial liquid-liquid supercooling transition -- as simple switchovers among the three bonding types.

\end{abstract}

\section*{Introduction}

Water is one of the most computer-simulated molecules in the world, because of its major roles in biology, chemistry, physics and geoscience.  Yet, it remains poorly understood.  Liquid water is unusual in having a density maximum in its liquid temperature range, a change in sign of its thermal expansion, minima in both its compressibility and heat capacity, and an apparent liquid-liquid critical point in its supercooled region.  We seek explanations that are rooted in the structure and energetics of individual water molecules.  There have been two general approaches to physical modeling of water’s structure property relationships.  \\

First, water has been simulated by computations of ever-improving atomistically detailed models, such as  ST2, TIP, SPC, OPC, rWAIL, and mW \cite{Rahman1971, Jorgensen1981, Abascal2005, Berendsen1987, Molinero2009, Izadi2014, Medders2014, Khalak2018, Weldon2024}, or polarizable models such as MB-pol, TTM, iAMOEBA, and MB-UCB \cite{Palos2024, Burnham2002, Wang2013, Reddy2016, Das2019, Lambros2020}, or machine learning (ML) and neural network (NN) models trained on $ab\ initio$ data \cite{ Morawietz2016, Singraber2018, Chan2019, Reinhardt2021, Wohlfahrt2020, Omranpour2024, Hijess2024, Liu2022, Bore2023, Zorzi2024}.  These approaches offer the most uncorrupted grounding in atomistic physics; typically quite good agreement with experimental data; and good transferability across different application venues.  But computer simulations have limitations:  (i) They are \textit{computationally expensive.}  As modeling challenges become ever bigger, especially in biology, today’s water modeling doesn’t scale well.  (ii) Simulations produce \textit{trajectories,} which are fluctuating vectors that become interpretable only upon converged averaging over multiple runs.  They have sampling errors, so computed heat capacities and other derivatives of free energies are challenging.  For these reasons, a major controversy took years -- and multiple forcefields -- to resolve; namely the nature of the supercooled liquid transition \cite{Poole1992, Limmer2011, Limmer2013, Palmer2014, Debenedetti2020}.   (iii) Simulations lack \textit{interpretability}.  They are snapshots, not explanations.  MD trajectories are strings of stochastic many-atom, many-bond, many-configuration microstates.  While such modeling often reproduces water’s properties, it doesn’t explain them. \\

A second approach is more coarse-grained, based on defining mesostates – such as hydrogen bonded, or in ice-like clusters, or having different local water densities, for example. This approach arguably began with Roentgen in 1892 \cite{Rontgen1892}, who interpreted liquid water as having two components of different densities.  Bernal and Fowler \cite{Bernal1933} identified the importance of hydrogen bonding in ice structure and Pauling \cite{Pauling1935} estimated the residual entropy of ice from hydrogen arrangements between the oxygen. Pople \cite{Pople1951} suggested the existence of a continuum of distorted hydrogen bonds. In contrast, Davis and Litovitz \cite{Davis1965} and Angell \cite{Angell1961} proposed two-state models, consisting of ice-like open hexagonal waters and closed-packed waters. Other two-state thermodynamic theories \cite{Poole1994, Truskett1999, Tanaka2000a, Bertrand2011, Holten2012} posit that water anomalies are a result of competition between a less ordered high-density liquid (HDL) and a more ordered low-density liquid (LDL) water states, reviewed in detail by Gallo \textit{et al.} \cite{Gallo2016}. In recent advances, the otherwise incommensurate tetrahedral and spherical symmetries can now be treated together within a single framework \cite{Urbic2018, Franzese2003, Franzese2007}. In the Franzese-Stanley model (FS) \cite{Franzese2003, Franzese2007}, water molecules are placed onto lattice sites and interact with neighboring waters through mean-field Ising contact and orientational interactions, in addition to a 3-body cooperative tetrahedrality energy term \cite{coronas2024p}. But the FS model relies on simulating Monte Carlo trajectories.  Here, we aim for a more analytical theory instead.  \\

We build upon a model of Urbic \cite{Urbic2018}, which is off-lattice, reckons with pairwise Lennard-Jones and pairwise hydrogen bond interactions and has a 12-body ice-like cooperativity energy.  But here we introduce translational and volumetric physics, replacing prior simplifications.  The present model, \emph{Cage Water}, starts from a partition function, so it is analytical, not sampled; thus orders of magnitude faster than a simulation.  The novelty of this model is in bridging from interpretable molecular physics on the one hand, to making excellent predictions of extensive thermophysical experiments on the other hand.  It also gives insights into a controversy about water's supercooled liquid-liquid transition.

\section*{Theory of the Three Bond Types}

\begin{figure}[!htb]
\centerline{\includegraphics[width=5in]{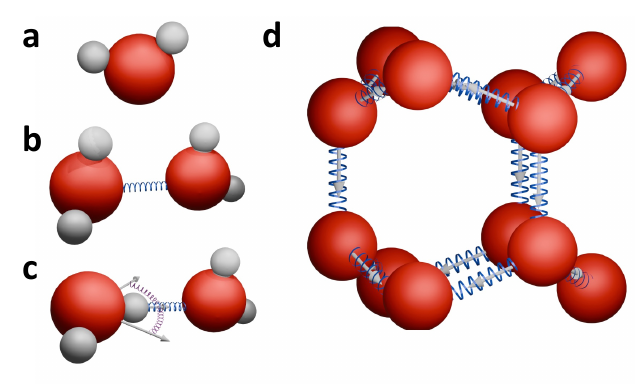}}
\caption[]{\label{fig: schematic}  \textbf{Cage Water's four  microstates.}   \textbf{a)} Non-interacting (NI) monomer water. b) Pair van der Waals (vdW) interaction. Translational spring in blue. \textbf{c)} Pair HB (pHB) state with HB arms shown with an arrow. Orientational springs (purple) and the HB arms (arrows) indicate dynamic HBs. \textbf{d)} A 12-member cooperative cage (cage) unit of Ih ice, showing connecting translational springs. a,b,c) Oxygen atoms are shown with red balls and hydrogen atoms are shown with silver balls. Hydrogen atoms are not shown in \textbf{d)} for clarity}
\end{figure}

We define four different mesostates of liquid water, figure \ref{fig: schematic}, similar to the Cage Water model \cite{Urbic2010}: i) isolated non-interacting water molecules, such as those in the vapor phase (NI); ii) water pairs that do not have a hydrogen bond but interact through a van der Waals (vdW) type interaction; (iii) water pairs that interact through a hydrogen bond (pHB); and iv) a fully hydrogen-bonded cooperative ice-like cage (cage) in the liquid state. Below, we define the statistical mechanical weights of these four states, cum them into a partition function, and solve for the various state populations as functions of temperature and pressure. \\ 

First, we estimate the volumes of four states from a reference water species, Ih ice. We model the liquid water cage state in the limit of the Ih ice cage, consisting of 12-membered building units of water cages. We geometrically estimated the equilibrium volume of the cage per water molecule, 
\begin{equation}{\label{eq:volume}}
    v_{cage}=\frac{8\sqrt{3}\sigma_{Ih}^3}{9} = A_{cage} \sigma_{Ih}^3,
\end{equation}

\noindent{where $\sigma_{Ih}$ is the water-water distance in the Ih ice cage. We perturb the cage water to model pHB state, $A_{pHB} =x_{pHB}\times A_{cage}$, and vdW state, $A_{vdW}=x_{vdW}\times A_{cage}$, and NI state in in the vapor limit, $v_{NI}=v_{vdW}+T/p$. We allow $\sigma_j$ of each state to contract or expand in response to the thermodynamic condition to capture the translational contributions. See the supplementary information (SI) for details.}\\

We model the direction HB interaction, equation \ref{eq:HB_eps}, with an attractive constant part $-\epsilon_{pHB}^0$, a restoring orientation-dependent harmonic potential with spring constant $\kappa_\Theta$ with equilibrium angle between arms $\Theta =0$  where the rotational penalty is $0$.
\begin{equation}{\label{eq:HB_eps}}
\epsilon_{pHB} = -\epsilon_{pHB}^0 + \kappa_{\Theta}(1-\cos(\Theta))^2\ ; \quad \quad for -\pi/6 <= \Theta <= \pi/6.   
\end{equation}
Interactions in cooperative cages differ from the HB interaction by an additional attractive term, $\epsilon_{cage} = \epsilon_{pHB} -\epsilon_c$. Whereas vdW state only consists of the constant attractive term,
\begin{equation}
\epsilon_{vdW} = -\epsilon_{vdW}^0. 
\end{equation}
We capture the translational contribution to energy due to the contraction and expansion of volumes with a restoring harmonic potential, $ \kappa_x \langle\Delta x\rangle^2$ with a spring constant $\kappa_x$ and with the same translational limits as volumes, see SI for details.\\

Following experimental observations and for simplicity, we assume that each water in the three states, cage, pHB, and vdW, interacts with four neighboring water molecules. There would be an additional $6k_BT$ to the overall energy per water molecule, which accounts for equally partitioned bond vibration energy for four bonds. We can estimate the effective interaction energy per water molecule as,
\begin{equation}{\label{eq:overall_u}}
u_j = -2\epsilon_j + \kappa_x \langle\Delta x\rangle^2 + 6k_BT   
\end{equation} 
\noindent{Interaction energy for NI state is zero i.e. $u_{NI} =0$.}\\

Now we estimate the Boltzmann weights for states, $\Delta_{cage}, \Delta_{pHB}, \Delta_{vdW}$, and $\Delta_{NI} $, with energy $u_j$ and volume $v_j$, in an isothermal-isobaric ($NpT$) ensemble by integrating over rotational degrees of freedom $\Phi, \Psi$, and $\Theta$ and translational degrees of freedom $x, y$, and $z$, see the SI for details. \\

From these Boltzmann weights, we estimate the partition function for the 12-membered single cage as
\begin{equation}
    Q_1=(\Delta_{pHB}+\Delta_{vdW}+\Delta_{NI} )^{12}-\Delta_{pHB}^{12}+\Delta_{cage}^{12}.
\end{equation}
Here, the first term accounts for many-body combinations, up to $12^{th}$ order, of the three water states, HB, vdW, and NI, contributing to 12 water molecules. We assume there is additional cooperative energy when all 12 waters in the cage are hydrogen bonded with each other. To capture this cooperativity, we replace the $12^{th}$ order pair HB contribution $\Delta_{pHB}^{12}$, the $2^{nd}$ term with $\Delta_{cage}^{12}$, the $3^{rd}$ term. The partition function for one water effectively would be given by $q_b=Q_1^{1/12}$.

We model the attraction between 12-membered cages with a self-consistent mean-field approximation similar to Truskett and Dill (TD) \cite{Truskett2002a}. The additional dispersion interaction between $N$ molecules is $-Na/\langle v_{mol}\rangle$, where $a$ is the van der Waals dispersion term between cages and $\langle v_{mol}\rangle$ is the molar volume of single water molecules. This dispersion term reduces the system's pressure from $p_0$ to $p = p_0-a/\langle v_{mol}\rangle^2$. So, to estimate properties at pressure $p$, we evaluate them at higher pressure $p_0$ such that
\begin{equation}
    p_0 = p+a/\langle v_{mol}\rangle^2.
\end{equation} 
We then obtain water's thermophysical properties using standard thermodynamic relationships \cite{Dill2011} from the partition function \cite{Truskett2002a, Urbic2010, Urbic2012, Urbic2018}.

\section*{How the model explains liquid water's properties}

\begin{figure}[!htb]
    \centering
    \includegraphics[width=6in]{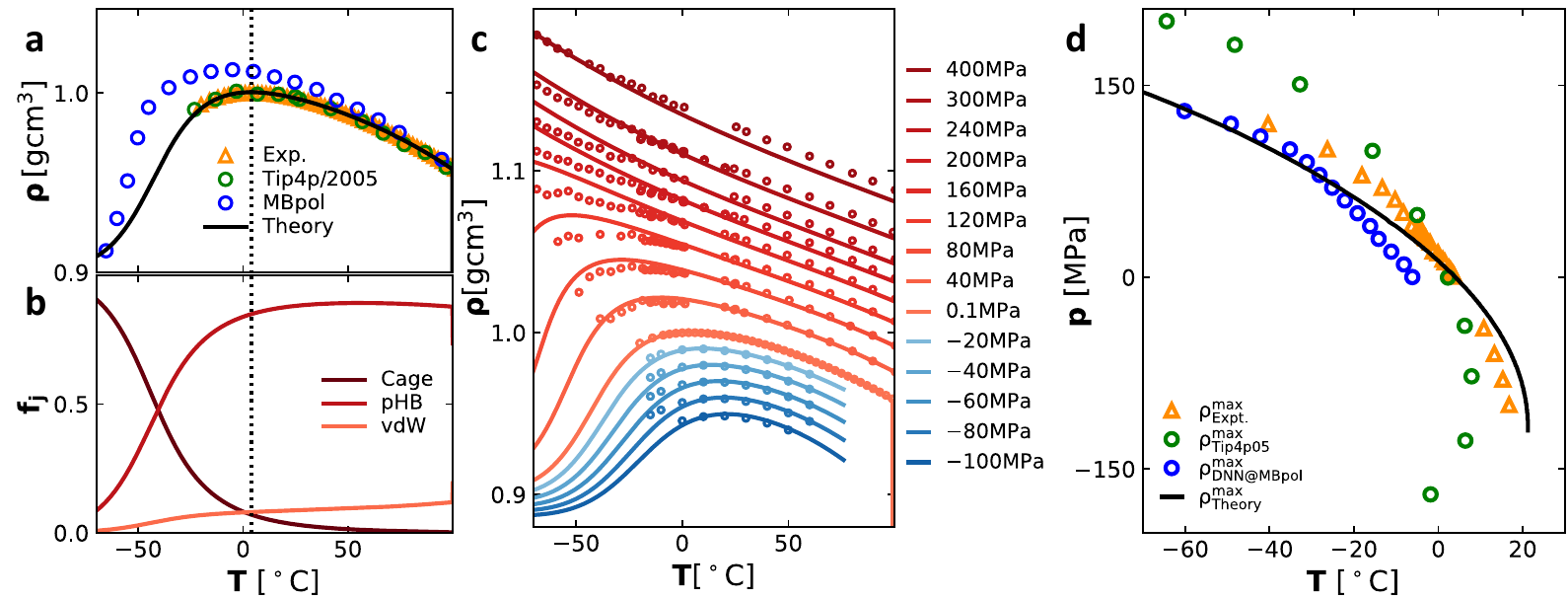}
    \caption{\textbf{Liquid water's density vs. temperature and pressure, predictions and experiments.}  \textbf{a)} Density vs. temperature at atmospheric pressure: model (solid black line); experiments\cite{Mishima2010, Pallares2016} (orange open triangles); TIP4P (open green circles); MB-pol (open blue circles). \textbf{b)} The computed populations of the three model microstates. The dotted line in \textbf{a, b} indicates the temperature of maximum density, 4 $^\circ$C. \textbf{c)} Density isobars for non-ambient pressures. Theory predictions for $p>0$ are shown in red and for $p<0$ in blue. Circles are the experimental data points. \textbf{d)} Maximum density prediction in comparison with experiments (orange triangles), TIP4P (green circles), and DNN@MBpol \cite{Sciortino2025} (blue circles). Pressures are in MPa.}
    \label{fig:density}
\end{figure}

A purpose of this model is to provide a microscopic interpretation of water's liquid across temperature and pressure, based on the calculated populations ($f_{cage}, f_{pHB}, \And f_{vdW}$) of the microstates, figure \ref{fig:density}b.  Here we compare the model predictions: (i) to experiments and (ii) to the TIP4P/2005 \cite{Abascal2005, Gonzalez2016} and to MB-pol \cite{Palos2024} explicit water models, which are often regarded as the state-of-the-art atomistic models for pure liquid water at least at ambient pressure.

\subsection*{Water's density, a balance of open and compact structures}
We compute water's density from the molar volume, $v_{mol}$, with typical density relation $\rho = M_w/v_{mol}N_A$, where $M_w = 18\ g/mol$ is molecular weight of water per mole and $N_A = 6.022 \times 10^{23}\ molecules/mol$ is the Avogadro number. Figure \ref{fig:density} shows that this model quantitatively predicts the density for the entire range $pT$ ($-100\ MPa\ to\ 400\ MPa \And -75^\circ C\ to\ 150^\circ C $) explored experimentally \cite{Mishima2010, Pallares2016}. At atmospheric pressure, the prediction is as accurate as that of TIP4P/2005 and MB-pol. It predicts a density maximum around 4$^\circ$C, shown with a vertical dotted line in figure \ref{fig:density}a, and also the boiling at 100$^\circ$C (see figure S1).\\

At temperatures colder than $\approx -45 ^\circ$C, where cage and pHB states crossover, the cage state dominates owing to its stronger cooperative energy, resulting in highly ordered ice-like low-density cagey water. At warmer temperatures, pHB water dominates, due to its smaller volume than cage water and stronger interactions than vdW water, causing density to increase until a crossover between vdW and cage state near $ 4 ^\circ$C. Beyond this point, vdW water population increases significantly due to its higher translational and rotational entropy, causing the density to decrease, resulting in a maximum at 4$^\circ$C. \\

Figure \ref{fig:density}c shows predicted density-temperature curves for different pressures. First, as expected, water's density increases with pressure. Figure S2 shows the model prediction that increasing pressure crunches cages into pHB pieces. Figure S3 shows that the two crossover points, from $f_{cage}$ to $f_{pHB}$ and from $f_{cage}$ to $f_{vdW}$, shift to lower temperatures with increased pressure, resulting in the temperature of maximum density (TMD) shifting towards a lower temperature with pressure.  Figure \ref{fig:density}d experimental predictions that are comparable to, or better than, those of TIP4P/2005 and the deep neural network surrogate of MB-pol (DNN@MB-pol) \cite{Sciortino2025}. Under stretchable pressures, $p<0$, TMD undergoes a turnaround point, resulting from the nonmonotonic behavior of $f_{vdW}$. It increases with pressure in cold water, $T<\approx 10^\circ$C until it plateaus at $\approx9\%$, figure S2c. \\

Two-state models \cite{Angell1961, Poole1994, Truskett1999, Tanaka2000a, Bertrand2011, Holten2012, Gallo2016, Caupin2019} treat the water anomalies as a competition between a less ordered high-density liquid (HDL), and a highly ordered low-density liquid (LDL). These models capture supercooled water properties quite well. However, the presence of highly ordered LDL at warmer temperatures is counterintuitive. Ricci et al. \cite{Ricci2008} found experimentally that this highly ordered LDL water predominantly exists at much colder temperatures. This suggests a presence of different low-density water that is less ordered, the vdW water in this model, and its population increases due to the breaking of more HBs at warmer temperatures, causing the density to decrease. This is also supported by the findings of Morawietz et al. \cite{Morawietz2016} where van der Waals interactions were crucial to reproducing density maximum in an $ab\ initio$ quality NN water potential. Under moderate stretchable pressure, $p<0$, more HB water are broken into vdW water due to tension. This increase in weak vdW population stabilizes water as it allows water to expand and thus nullifies the tension, figure S2c.\\

\begin{figure}[!htb]
    \centering
    \includegraphics[width=4in]{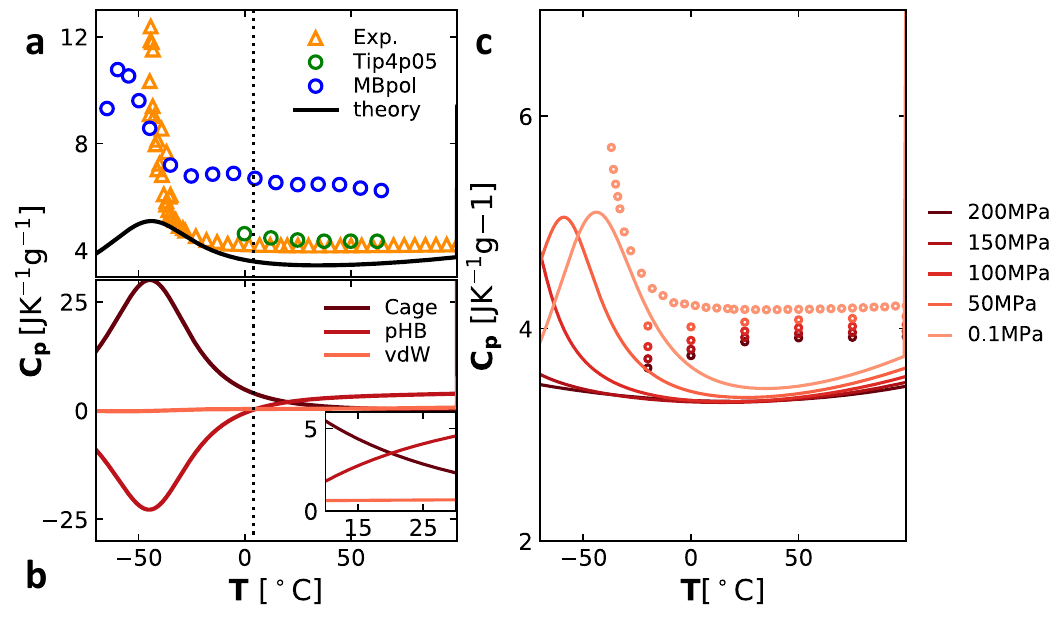}
    \caption{\textbf{Heat capacity, $C_p$, and its components vs temperature.}  \textbf{a)} $C_p$ at atmospheric pressure. \textbf{b)} Model component contributions to $C_p$. \textbf{c)} $C_p$ isobars. Pressures are in MPa. Legends are the same as in figure \ref{fig:density}. Experimental data is taken from ref. \cite{Angell1982, Mallamace2014, Lin2012, pathak2021}.}
    \label{fig:cp}
\end{figure}

\subsection*{Heat capacity minimum bonding transition: cage HB $\rightarrow$ pHB}

We calculate the heat capacity from the enthalpy using the thermodynamic relationship 
\begin{equation}
    C_p = \frac{\partial  \langle h \rangle}{\partial T} =\sum_j \frac{\partial  (f_j \langle h_{j} \rangle)}{\partial T} \equiv \sum_j C_{p,j},
\end{equation}
here $C_{p,j}$ is the contribution from the $j^{th}$ state. Figure \ref{fig:cp} shows agreement with experiments no worse than TIP4P/2005 or MB-pol at ambient temperatures. It also captures the anomalous increase in the supercooled region. The pronounced minimum (compared to a subtle minimum observed in experiments \cite{Debenedetti2003}) is around $35^\circ$C, which results from the dominant contribution switch from cage to pHB, see figure \ref{fig:cp}b. The maximum deviation from the experiment is near this minimum, $\approx 0.6\ J/gK$, which is reasonable. At lower temperatures, the contribution from breaking of cages dominates which causes $C_p$ to increase until it peaks, $C_p^{max}$, around $-44 ^\circ$C, figures \ref{fig:cp}b. Confined water experiments \cite{Mallamace2008, Mallamace2014, Oguni2008} as well as theories (MD simulations \cite{Biddle2017, Gonzalez2016}, equation of state models \cite{Biddle2017, Holten2012, Holten2014b}, and recent statmech based model \cite{coronas2024p}) exhibit such peaks. At high pressure, figure \ref{fig:cp}c shows that predicted $C_p$, similar to experiment, decreases with pressure at ambient temperature and $C_p^{max}$ shifts to a lower temperature (see figure S4) similar to TMD. This is a consequence of the shift in the crossover point between the cage and pHB water to a lower temperature with pressure, figure S3.

\begin{figure}[!htb]
    \centering
    \includegraphics[width=4in]{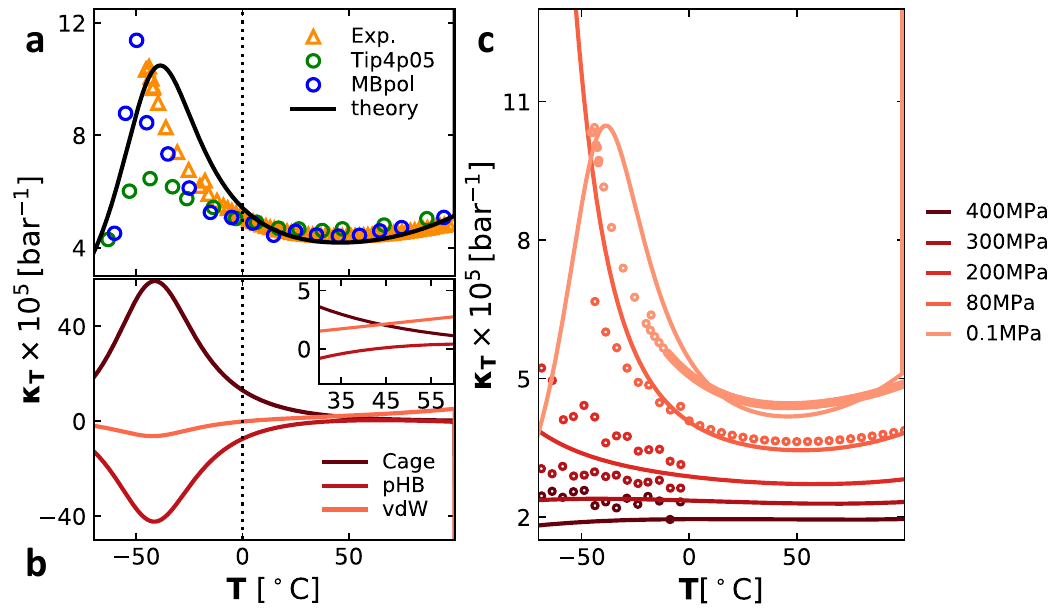}
    \caption{\textbf{Isothermal compressibility, $\kappa_T$, and its components vs temperature.} \textbf{a)} $\kappa_T$ at atmospheric pressure. \textbf{b)} Microscopic components of $\kappa_T$. The inset zooms in at the cage  $\rightarrow$ vdW crossover. \textbf{c)} $\kappa_T$ isobars. Pressures are in MPa. Legends are the same as in figure \ref{fig:density}. Experimental data is taken from ref. \cite{Mishima2010, Chen1977, Kim2017}.}
    \label{fig:kappa}
\end{figure}

\subsection*{Compressibility minimum bonding transition: cage HB $\rightarrow$ vdW}

We calculate the isothermal compressibility as the volumetric fluctuations in the molar volumes, $\langle v_{mol}\rangle$ with respect to pressure as, 
\begin{equation}
\begin{split}
    \kappa_T = -\frac{\partial \ln \langle v_{mol}\rangle}{\partial p} = -\frac{1}{\langle v_{mol} \rangle}\ \sum_j \frac{\partial f_j \langle v_j \rangle}{\partial p} \equiv \sum_j \kappa_{T,j},    
\end{split}
\end{equation}
where $\kappa_{T,j}$ is the compressibility of state type $j$. Figure \ref{fig:kappa} shows that the model predictions are as good as TIP4P/2005 
and at 1 atm at ambient conditions and captures much broader experimental data. The anomalous minimum, $\kappa_T^{min}$, near $\approx 45 ^\circ$C is understood through the model as a switchover the dominance of cages to vdW waters; see the inset of figure \ref{fig:kappa}b. Below this temperature, the increase in $\kappa_T$ is due to an increase in the dominant contribution from cage waters, more compressible than pHB waters, which also peaks near the $\kappa_T^{max}$ at $\approx -40\ ^\circ$C in figure \ref{fig:kappa}a, in excellent agreement with the observed value, $\kappa_T^{max}$ at $\approx -44\ ^\circ$C  \cite{Kim2017}. In addition, sound velocities \cite{Holten2017} and confined-water experiments \cite{Mallamace2013} as well as computational and theoretical studies \cite{Abascal2010, Holten2012, Ni2016, Palos2024} also report $\kappa_T^{max}$ in supercooled water. Figure \ref{fig:kappa}c shows that $\kappa_T$ decreases with pressure, and $\kappa_T^{max}$ shifts to a lower temperature with pressure (see figure S4).

\begin{figure}[!htb]
    \centering
    \includegraphics[width=4in]{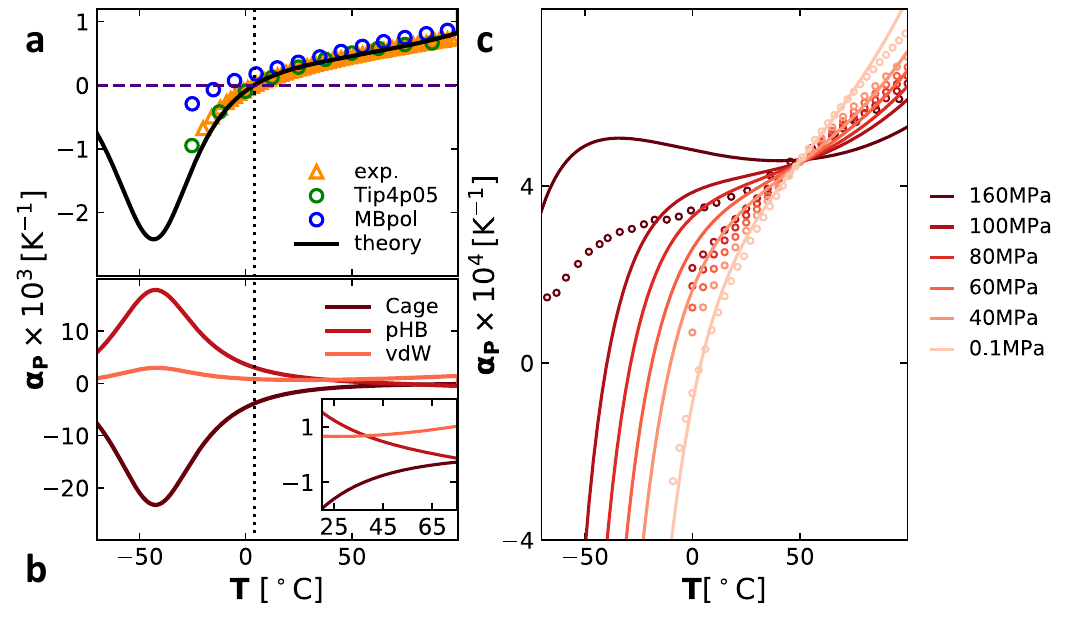}
    \caption{\textbf{Thermal expansion coefficient, $\alpha_p$, and its components vs temperature.} \textbf{a)} $\alpha_p$ at atmospheric pressure. \textbf{b)} Microscopic components of $\alpha_p$. The inset zooms in at the pHB $\rightarrow$ vdW and pHB $\rightarrow$ cage crossovers. \textbf{c)} $\alpha_p$ isobars. Legends are the same as in figure \ref{fig:density}. Experimental data is taken from ref. \cite{Mallamace2014, Chen1977}.}
    \label{fig:alpha}
\end{figure}

\subsection*{Thermal expansion bonding transition: pHB $\rightarrow$ vdW}
We calculate the expansion coefficient as the volumetric fluctuations in the molar volumes, $\langle v_{mol}\rangle$ with respect to temperature, 
\begin{equation}
\begin{split}
    \alpha_p = \frac{\partial \ln \langle v_{mol}\rangle}{\partial T} = \frac{1}{\langle v_{mol} \rangle}\ \sum_j \frac{\partial f_j \langle v_j \rangle}{\partial T} \equiv \sum_j \alpha_{P,j},    
\end{split}
\end{equation}

where $\alpha_{p,j}$ is the expansion coefficient of the $j^{th}$ state. Figure \ref{fig:alpha} shows that the current model captures the experimentally observed expansion coefficient for a broad $pT$ range. The abnormal negative $\alpha_p$ below $4 ^\circ$C is the result of the dominant negative contribution from cage water, figure \ref{fig:alpha}b. This reflects the loss of water's expandability as the more expandable pHB waters transition into less expandable cage waters with stronger and rigid HBs. \\ 

Similar to $C_p^{max}$ and $\kappa_T^{max}$, $\alpha_p$ exhibits a minimum, $\alpha_p^{min}$, in the supercooled region (see figure S4). However, in the normal temperature range, $\alpha_p$ isobars are nonmonotonic and crossover at $\approx 50^\circ$C, an inflection point where water stops behaving abnormally \cite{Mallamace2014}. This model predicts a inflection near $\approx 48^\circ$C. Figures \ref{fig:alpha}b and S3b shows a transition from pHB to vdW in the vicinity of the inflection point, suggesting more vdW liquid like expansion behavior beyond this crossover. Additionally, figure S2c shows that the change in $f_{vdW}$ behavior beyond this inflection point is the cause behind the $\alpha_p$ isobars crossover around $\approx 50^\circ$C.\\

\subsection*{Supercooling liquid transition bonding transition: cage HB $\rightarrow$ pHB.}

\begin{figure}[!htb]
\centerline{\includegraphics[width=6in]{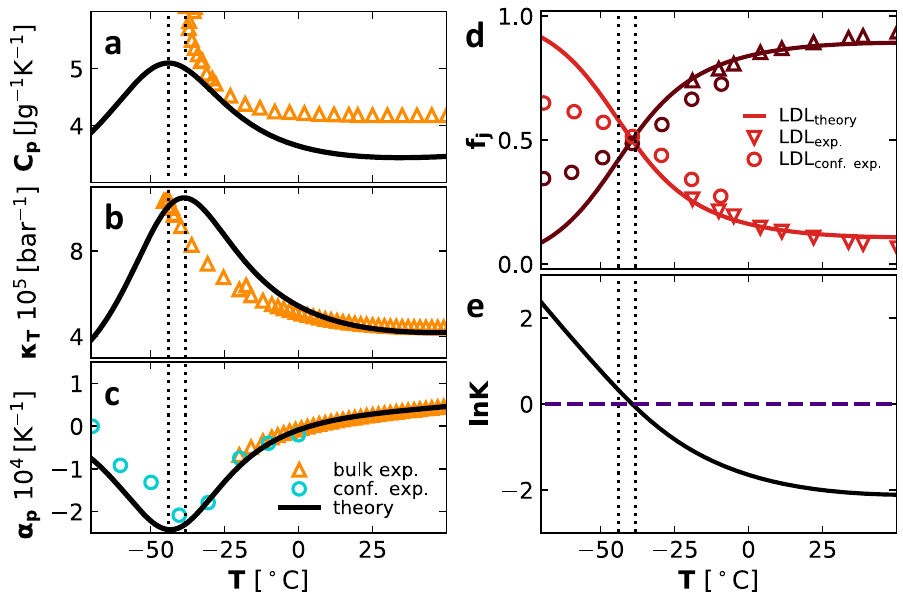}}
\caption[]{\label{fig: all_prop}  \textbf{Extrema in supercooled water at 1 atm are near HDL $\rightleftharpoons$ LDL equilibrium.}   \textbf{a)} $C_p$, partial experimental data is shown to emphasize the predicted peak. \textbf{b)} $\kappa_T$. and \textbf{c)} $\alpha_p$. \textbf{a)-c)} The model predictions (black solid lines), bulk water experiments \cite{Angell1982}(orange triangles), and confined water experiments \cite{Mallamace2014}(cyan triangles). \textbf{d)} Population of HDL ($f_{HDL}$) in brown and LDL ($f_{LDL}$) in red. 
Theory predictions are shown with solid lines, bulk experimental estimates of $f_{LDL}$ taken from \cite{Tyburski2024} and $f_{HDL} = 1-f_{LDL}$ shown with open triangles, and confined experimental estimates are taken from  \cite{Mallamace2009} shown with open circles. \textbf{e)} $\ln K$ with vs temperature. $K=f_{LDL}/f_{HDL}$ is the equilibrium constant for the two-state system, HDL $\rightleftharpoons$ LDL.} 
\end{figure}

There has been controversy about supercooled water.  Is there a liquid-liquid phase transition?  Support for such a transition is shown in Figure ~\ref{fig: all_prop}, indicating a tendency towards an extremum in the heat capacity, compressibility, and expansivity upon lowering the temperature to around -44 $^\circ$C \cite{Kim2017, pathak2021}. Such extrema are well-established indicators of a phase change. In 1992, Poole et al. \cite{Poole1992} proposed that such peaks in $C_p$ and $\kappa_T$ indicate the existence of Liquid--Liquid Transition (LLT) line between a high density liquid (HDL) of some kind and low density liquid (LDL); it meets the locus of maximum correlation length in the single phase, the Widom Line (WL), at the Liquid--Liquid Critical Point (LLCP), known as the LLCP scenario. Subsequently, evidence has been sought in molecular simulations. 
Several atomistic simulations using ST2 models \cite{Poole1992, Poole2005, Liu2009, Cuthbertson2011} and TIP models \cite{Yamada2002, Paschek2005, Paschek2008, Abascal2010, Corradini2010} indicated the LLCP in the so-called \emph{no man's land,} where water readily crystallizes into ice.  In contrast, the mW model doesn't show an LLCP \cite{Molinero2009}. Limmer and Chandler argued that these MD simulations suffer from inadequate sampling and dependencies on forcefields, and that the transition being observed in the simulations was simply water's crystallization, But subsequently, Palmer et al. \cite{Palmer2014} showed the existence of two metastable liquid phases, LDL and HDL, in equilibrium with a crystalline state for the ST2 model at 228.6 K and 240 MPa, independent of different advanced sampling methods. \\

More recent modeling and thermodynamic analyses have indicated the presence of LLT and LLCP points for supercooled water \cite{ Ni2016, Debenedetti2020, Gartner2022b, Espinosa2023, Weldon2024, Sciortino2025, dhabal2024, Coronas2024a, Caupin2019, Mishima2023}. Direct experimental observation of the LLT and LLCP have been elusive until recently.  New experiments in ultrafast X-ray spectroscopic techniques have recently accessed the \emph{no man's land} region of water's supercooled phase diagram \cite{Sellberg2014,Kim2017, Kim2020, pathak2021, Nilsson2022, Amann-Winkel2023}. Kim et al. \cite{Kim2020} provided direct experimental evidence of structural changes between HDL and LDL phases at 205 $\pm$ 10 K and pressure between 0.1 - 350 MPa on isochoric heating of high density amorphous ice (HDA), indicating an LLT. A similar experiment showed structural changes between HDL and LDL phases at 200 K on isochoric heating of low density amorphous ice (LDA) \cite{Amann-Winkel2023}, further supporting the existence of LLT and LLCP scenario. However, the  big range of estimates of pressure (from ambient to 350 MPa \cite{Kim2020, Amann-Winkel2023}) imply that direct observation of LLCP through experiments is still out of reach. \\

How does our Cage Water modeling add to this story?  First, the challenge faced by all the MD studies is that they are simulations, requiring converged averages over trajectories that require extensive sampling of small free energy differences at very cold temperatures, where simulations are notoriously difficult \cite{Holten2014b, Palmer2014, Pallares2014, Abascal2010, Wikfeldt2011, Santra2015, English2011, Mallamace2013, Ni2016, Smallenburg2015, Debenedetti2020, Sciortino2025}. To converge on a story has required tests in multiple forcefields, and multiple long simulations.  In contrast, the present model is analytical, with no concern about trajectories, sampling, small barriers or cold temperatures.  The model is validated by good agreement with extensive liquid temperature and pressure experiments.  Most importantly, the present model reveals a story that is expressed in meaningful subpopulations of water's forms of bonding.  Some of the main players in the story -- particularly the caging clusters -- ``punch above their weight'', having importance for liquid state properties that is outsized compared to their very small populations. Details are below.\\

First, figure \ref{fig: all_prop} a--c shows that the Cage Water model predicts the same telltale divergence features as experiments show of the supercooling anomaly, which occurs between -44 to -38 $^{\circ}$C at atmospheric pressure. Second, while no bulk experiments have yet been able to reach temperatures low enough to probe the supercooling peak, experiments on nano-confined liquids can do so \cite{Mallamace2009}; see the blue circles in Figure \ref{fig: all_prop} c.  The Cage Water model predicts that the supercooling peak point is precisely where the liquid undergoes a transition from a low-density liquid at lower temperatures to a high-density liquid at higher temperatures.  Figure \ref{fig: all_prop} e shows the experimentally observed crossover point \cite{Tyburski2024}.  And Figures \ref{fig: all_prop} d and e show this experimental transition point to be where the model has a crossover from a dominant population of LDL (cage plus vdW) to a dominant population of HDL (pHB states). \\

\begin{figure}[!htb]
\centerline{\includegraphics[width=3in]{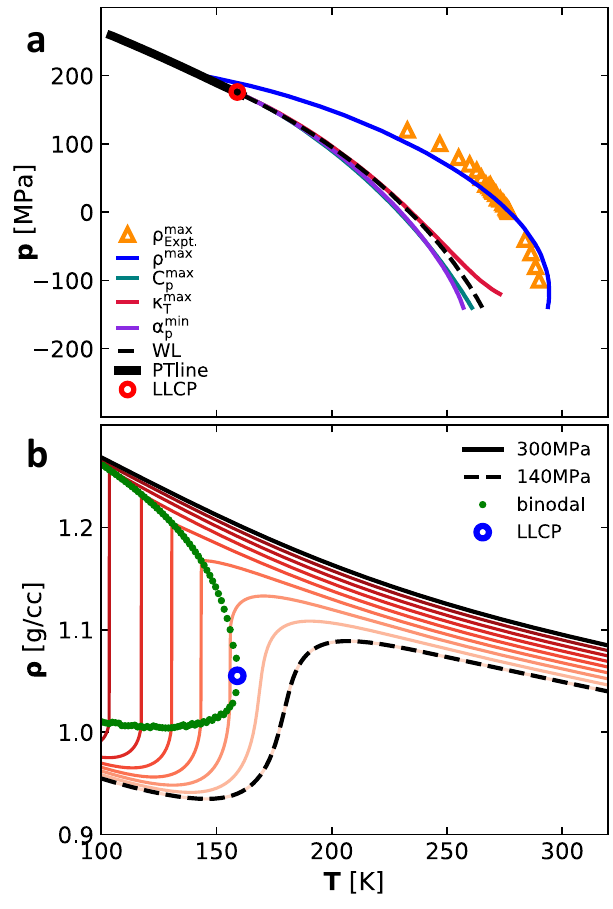}}
\caption[]{\label{fig: PT_phase} \textbf{The model captures the pressure dependence of the supercooling peak temperatures.} \textbf{a)} Model predictions of response function extrema are shown with solid lines; $\rho^{max}$ in blue solid line, $C_p^{max}$ in teal, $\kappa_T^{max}$ in crimson, and $\alpha_p^{min}$ in violet. The predicted Widom line (WL) and phase transition (PT) line are shown with a black dashed line and a thick solid line, respectively. Experiments \cite{Pallares2016, Mishima2010, Sotani2000, Caldwell1978} are shown with open orange triangles. \textbf{b)} Density isobars for a range of 140-300 MPa on an interval of 20 MPa, showing first-order phase transitions, are shown in shades of red. The phase boundary, the binodal line, is shown with green dots. The predicted liquid-liquid critical point at pressure 176 MPa and temperature 159 K is shown with an open red circle in panel  \textbf{a)} and an open blue circle in panel \textbf{b)}.}
\end{figure}

Figure~\ref{fig: PT_phase} shows that Cage Water captures well the thermophysics of water's supercooling.  First, some basic expectations: The Liquid-Liquid Critical Point (LLCP) should be the point on the PT phase diagram where the Liquid-Liquid Phase Transition (LLPT) ends and the Widom Line (WL) begins (the loci of the maximum correlation lengths) and also where the response functions have their extrema ($C_p^{max}, \kappa_T^{max}$, and $\alpha_p^{min}$) \cite{Franzese2007, gallo2014}. If a system is two-state (in our case the high- and low-density phases, HDL $\rightleftharpoons$ LDL), then the LLCP point, the LLPT line, and the WL line should all be where $\ln K = 0$, where $K$ is the equilibrium point of the transition \cite{Singh2016}. Figure~\ref{fig: PT_phase} shows this holds true for Cage Water. The model predicts the LLCP to be at 159 K and 176 MPa, figure \ref{fig: PT_phase}a. Figure \ref{fig: PT_phase}b shows the predicted transition curves, the critical point, the binodal curve, and the first-order nature of the transition below the critical temperature. As far as we know, there are no experimental measurements of the critical point yet.  But our predicted critical pressure is within the expected range of $~143-360$ MPa \cite{Poole1992, Mishima1998, Poole2005, Liu2009, Cuthbertson2011, Yamada2002, Paschek2005, Paschek2008, Abascal2010, Corradini2010, Palmer2014, Debenedetti2020, Gartner2022}.  And $\rho^{max}$ is in excellent agreement with the experiments.  However, the model critical temperature is somewhat colder than the previously predicted range of $~172-250$ K.

\subsection*{The model energies are consistent with known values} 

The predicted water-water bond (dissociation) energy, $\langle u \rangle = \sum f_j u_j$, is temperature dependent but at room temperature, the model value is $\approx -12.7\ kJ/mol$(see figure S5). This is fairly consistent with different experiments: older thermal conductivity and infrared absorption data gives is $\approx -18\pm3 kJ/mol$ \cite{Curtiss1979, Gebbie1969, Dianov-Klokov1981, Bondarenko1991, Jin2003, Cormier2005}, while more recent estimates at room temperature using pressure broadening give $-13.6\pm4 kJ/mol$ \cite{Nakayama2007} and $-15.5\pm5.6\ kJ/mol$ \cite{Fiadzomor2008}, and velocity map imaging gives the $(H_2O)_2$ dissociation energy as $-13.2\pm0.12\ kJ/mol$ \cite{Blithe2011}.  Within similarly large ranges, these values are consistent with the corresponding $ab\ initio$ dissociation energy estimates of $D_0 \approx -13.2\ kJ/mol$ for $(H_2O)_2$ \cite{Shank2009}.  Experiments, however, don't give the microscopic components (vdW, pair HB, and cooperative). The values of these quantities in the model are: maximum strength (in the absence of thermal contributions) of a vdW energy is $\epsilon \approx -10.3\ kJ/mol$, and pHB and cage water at the minimum energy configuration of HB, when the HB arms are aligned are $\epsilon_{pHB}^0 \approx-19.2\ kJ/mol$ and $\epsilon_{cage}^0 \approx-20.5\ kJ/mol$), respectively (figure S5b) which are also consistent with corresponding $ab\ initio$ electronic dissociation energy $D_e \approx 21\ kJ/mol$ \cite{Shank2009}.

\section*{Conclusions}
We describe here Cage Water, a statistical mechanical model of liquid water. It treats water's orientational and translational freedom and explains the liquid thermophysical properties and anomalies in terms of three interaction types: water-water vdW bonds of strength $\approx -10.3\ kJ/mol$, water-water Hbonds of strength $\approx-19.2\ kJ/mol$, and multi-water cage-structure Hbonds of strength $\approx-20.5\ kJ/mol$. In comparison with extensive temperature and pressure experiments, the Cage Water model is as accurate as explicit models TIP4P/2005 and MB-pol (for the pure liquid), but it has the advantages that it is analytical, not computational, and not requiring trajectories or sampling.  So it is orders of magnitude faster to compute.  \\

Most importantly, the Cage Water model gives a remarkably simple interpretation of water's anomalies, as transitions among the three bond types.  Ice melts when small populations of cages convert to small populations of vdW bonds in a large background sea of pairwise Hbonds.  Water's supercooling liquid-liquid transition happens when water cages melt into pairwise Hbonds in a small background sea of vdW bonding. The minimum in thermal compressibility occurs when cages convert to vdW bonds.  The minimum in the heat capacity results from the conversion of cages to pHB. The crossover from negative to positive thermal expansion in cold water is a transition from pHB to vdW states.

\section*{Acknowledgments}
We acknowledge the funding from NIH Grant RM1 GM135136 and the Laufer Center for Physical and Quantitative Biology at Stony Brook University. We thank Tomaz Urbic for fruitful discussions and for sharing the code for the CageWater (UD) model.


\newpage
\bibliography{Main_Text.bbl}

\end{document}


\title{\textbf{Supplementary Information: Statistical mechanical theory of liquid water}}
\author[1]{Lakshmanji Verma \thanks{Contributing author: lakshmanji.verma@stonybrook.edu}}
\author[1,2]{Ken A Dill \thanks{Corresponding author: dill@laufercenter.org}}
\affil[1]{\textit{{Laufer Center for Physical and Quantitative Biology}, {Stony Brook University}, {NY}, {USA}}}
\affil[2]{\textit{{Department of Physics $\And$ Department of Chemistry}, {Stony Brook University}, {NY}, {USA}}}

\maketitle

\section*{Theory}

We model the liquid water cage state in the limit of the Ih ice cage, which consists of 12-membered building units of water cages. We geometrically estimated the equilibrium volume of the cage per water molecule, 
\begin{equation}
    v_{cage}=\frac{8\sqrt{3}\sigma_{Ih}^3}{9} = A_{cage} \sigma_{Ih}^3,
\end{equation}
 \noindent{where $\sigma_{Ih}$ is the water-water distance in the Ih ice cage. To mimic the pressure response of water in cagey liquid, we evaluate the average volume, $\langle v_{cage} \rangle$ at every temperature and pressure. Then we model the liquid HB state and vdW state as a perturbation of the cage water, such that $A_{pHB} =x_{pHB}\times A_{cage}$ and $A_{vdW}=x_{vdW}\times A_{cage}$ where $x_{pHB}=0.73$ and $x_{vdW}=1.21$. The volumes of these three states relate as $\langle v_{vdW}\rangle>v_{cage}>v_{pHB}$. For the NI state, we estimate $v_{NI}=v_{vdW}+T/p$.} 

To capture the expansibility and compressibility of water over the $pT$ range, we assume the volume of each state is Boltzmann distributed and can be estimated as  
\begin{equation}
    \langle v_j \rangle = \frac{\int_{r_l}^{r_u^j} v_j \exp(-\frac{pv_j}{k_BT})r^2dr}{ \int_{r_l}^{r_u^j} \exp(-\frac{pv_j}{k_BT})r^2dr} \\
\end{equation}

Here $r_l$ and $r^j_u$ are fractional fluctuation limits in $\sigma_{Ih}$. The lower limit $r_l = \sigma_{Ih}(1-r_d)$ corresponds to the hard core (inaccessible core) of the water molecule, which is the same for all four states. Whereas the upper limit $r_u^j = \sigma_{Ih}(1+r_j)$, the expansion limit, differs for each state. For NI, we assume that the local volume of each state is $ \langle v_{vdW} \rangle$ and it can translate in the limit $0<r<\infty$ because $r_u^O>>r_l$ which reduces the integral to $\frac{k_BT}{p}$ and average volume per water molecule, $\langle v_{NI} \rangle$, to $ \langle v_{vdW} \rangle +\frac{k_BT}{p}$. \\

The direction HB interaction, equation (\ref{eqS:HB_eps}), is modeled with attraction $-\epsilon_{pHB}^0$ that corresponds to minimum energy configuration when HBs are aligned, a restoring orientation-dependent harmonic potential with spring constant $\kappa_\theta$ and reference orientation $\Theta_0 =0$ or $\cos(\Theta_0) =1$.
\begin{equation}{\label{eqS:HB_eps}}
\epsilon_{pHB} = -\epsilon_{pHB}^0 + \kappa_{\Theta}(1-\cos(\Theta))^2\ ; \quad \quad for -\pi/6 <= \Theta <= \pi/6.   
\end{equation}
Interactions in cooperative cages differ from the HB interaction by an additional attractive term, $\epsilon_{cage} = \epsilon_{pHB} -\epsilon_c$. Whereas vdW state only consists of an isotropic attractive term,
\begin{equation}
\epsilon_{vdW} = -\epsilon_{vdW}^0. 
\end{equation}
The interactions would weaken as the water molecules translate relative to each other on compression and expansion due to changes in $pT$ conditions. We estimate this loss of interaction as the penalty by modeling each interaction with a restoring harmonic potential with a spring constant $\kappa_x$ and equilibrium water-water distance $\sigma_{Ih}$. The Boltzmann averaged spring deformation, $\langle \Delta x\rangle$, over the same limit as for volumes is given by, 
\begin{equation}
    \langle\Delta x\rangle = \frac{\int_{r_l}^{r_u^j}  \Delta x\ \exp(-\frac{\kappa_x \Delta x^2}{k_BT}) dx}{\int_{r_l}^{r_u^j} \exp(-\frac{\kappa_x \Delta x^2}{k_BT})dx}. 
\end{equation} \\

For simplicity, we consider each water in the three states, cage, pHB, and vdW, interacts with four neighboring water molecules. There would be an additional $6k_BT$ to the overall energy per water molecule, which accounts for equally partitioned bond vibration. Therefore, we can estimate the overall energy per water molecule as 
\begin{equation}{\label{eqS:overall_u}}
u_j = -2\epsilon_j + \kappa_x \langle\Delta x\rangle^2 + 6k_BT   
\end{equation}
\noindent{Overall energy for open or non-interacting ($NI$) waters is zero i.e. $u_{NI} =0$.} \\

Now we can write the weight for single water in each state for an isothermal-isobaric ($NpT$) ensemble with energy $ u_j$ and volume $v_j$  per water molecule as 
\begin{equation}
    \Delta_j = c(T)4\pi\int_{r_l}^{r_u^j} r^2dr \exp\left(-\frac{p\langle v_j\rangle}{kT}\right) \int\int\int d\Phi d\Psi d\Theta  \exp\left(-\frac{\langle u_j\rangle}{kT}\right)
\end{equation}
$c(T)$ is the kinetic contributions to the weight which are assumed to be independent of the water state, $4\pi \int r^2dr$  integrates over the translational degrees of freedom, and  $\int\int\int d\Phi d\Psi d\Theta$ integrates over the rotational degrees of freedom. 

The weight for the HB state is given by 
\begin{equation}{\label{eqS:wgt_pHB}}
\begin{split}
    \Delta_{pHB} = c(T)4\pi\int_{r_d}^{r_{pHB}} r^2 \exp\left(-\frac{pv_{pHB}}{k_BT}\right)dr \times 4\pi^2\ \exp\left(-\frac{-2\epsilon_{pHB}^0-\kappa_x \langle\Delta x\rangle_{pHB}^2}{k_BT}\right) \\
    \times \int^{\pi/6}_{-\pi/6}   \ \exp\left(-\frac{2\kappa_{\Theta}(1-\cos(\Theta))^2}{k_BT}\right)sin(\Theta)d\Theta \\    
    \\
    \Delta_{pHB} = c(T)16\pi^3\ \exp\left(-\frac{-\epsilon_{pHB}^0 + \kappa_x \langle\Delta x\rangle_{pHB}^2)}{k_BT}\right) 
    \times \frac{-k_BT}{3pA_{pHB}} \exp\left(-\frac{pA_{pHB}r^3}{k_BT}\right)\bigg|_{r_d}^{r_{pHB}} \\
    \times \sqrt{\left(\frac{\pi k_BT}{8\kappa_{\Theta}}\right)} erf\left[ (2-\sqrt{3})\sqrt{\frac{\kappa_{\Theta}}{2k_BT}}\right]
\end{split}
\end{equation}
where $r_d$ is the inaccessible volume of water. $\Delta_{cage}$ is the same as $\Delta_{pHB}$ with an additional term $\exp(-\frac{2\epsilon_c}{k_BT})$ for cooperativity, the limit $r_{pHB}$ replaced with $r_{Cage}$, and volume coefficient $A_{pHB}$ replaced with $A_{cage}$. $\Delta_{vdW}$ state follows from equation (\ref{eqS:wgt_pHB}) without the rotational term, 
\begin{equation}{\label{eqS:wgt_vdw}}
    \Delta_{vdW} = c(T)\times32\pi^3\ \exp\left(-\frac{2(-\epsilon_{vdW}^0 + \kappa_x \langle\Delta x\rangle_{vdW}^2)}{k_BT} \right) 
    \times \frac{-k_BT}{3pA_{vdW}} \exp\left(-\frac{pA_{vdW}r^3}{k_BT}\right)\bigg|_{r_d}^{r_{vdW}}. \\
\end{equation}

In the absence of any interaction and in the translation limit of $0<r<\infty$, equation (\ref{eqS:wgt_vdw}) further reduces to 
\begin{equation}{\label{eqS:wgt_ni}}
    \Delta_{NI} = c(T)\times \frac{32\pi^3k_BT}{3pA_{NI}}. \\
\end{equation}
For the vapor weight calculation purpose, we have taken $A_{NI} = x_{NI}\times A_{cage}$, where $x_{NI} \approx 1700$, a physically reasonable expansion of water on vaporization.

From these weights for four states of water, we can write the partition function for the 12-membered single cage as
\begin{equation}
    Q_1=(\Delta_{pHB}+\Delta_{vdW}+\Delta_{NI} )^{12}-\Delta_{pHB}^{12}+\Delta_{cage}^{12}.
\end{equation}
Here, the first term accounts for many-body combinations, up to $12^{th}$ order, of the three water states, pHB, vdW, and NI, contributing to the 12-membered water. We assume that perfectly aligned four hydrogen bonds in the 12-membered cage impart an additional cooperative energy to each water. So, we replace the $12^{th}$ order pair HB contribution $\Delta_{pHB}^{12}$, $2^{nd}$ term, from the partition function, $Q_1$,  with $\Delta_{cage}^{12}$, $3^{rd}$ term to incorporate this cooperative nature of the 12-membered cage. The Partition function for one water effectively would be given by $q_b=Q_1^{1/12}$. Therefore, for a system of $N$ particles, the total partition function would be $Q=q_b^N=Q_1^{(N/3)}$, since each water is part of 4 cages. We then obtain water's thermophysical properties using standard thermodynamic relationships \cite{Dill2011} derived in prior works \cite{Truskett2002a, Urbic2010, Urbic2012, Urbic2018}.

Attraction energy between cages, beyond pairwise interactions, is modeled with a self-consistent mean-field approach, the same as in the TD model \cite{Truskett2002a}, $-a/\langle v_{mol}\rangle$ per water molecule, where $a$ is the van der Waal type dispersion term between cages and $\langle v_{mol}\rangle$ is the ensembles averaged molar volume of water, at given $T$ and $p$, given by 
\begin{equation}
    \langle v_{mol}\rangle = \sum_j f_j\langle v_j \rangle,
\end{equation}
where $f_j$ is the population of the respective state given by
\begin{equation}
    f_j = \frac{\partial \ln Q_1}{\partial \ln\Delta_j^{12}} = \frac{\Delta_j \partial  Q_1}{12 Q_1\partial \Delta_j}.
\end{equation}

\subsection*{Thermodynamics quantities after turning the mean-field attraction between cages on}
It can be shown that turning on the attraction between cages reduces the pressure of the system from $p_0$ to 
\begin{equation}
    p = p_0-a/ v_{0}^2,
\end{equation}
where $p_0$ is the pressure exerted by cages without the attraction and $p$ is the effective pressure of the bulk water. Similarly, it can also be shown the the enthalpy reduces from $h$ at pressure $p$ to $h_0 - 2a/v_0$ at pressure $p_0$ \cite{Truskett2002a}.

So, to get the properties at pressure $p$, we first evaluate them at higher pressure $p_0$ such that
\begin{equation}\label{meanfP}
    p_0 = p+a/ v_{0}^2.
\end{equation} 
We start from the actual thermodynamic pressure $p$ to self-consistently evaluate $p_0$ using Eq. \ref{meanfP}. Then the water properties at $p$ are calculated using relations from the properties evaluated at $p_0$. These relations, shown below, can be derived thermodynamically. Note that $v_{0}$ is a function of $p$, $p_0$, and $T$.

The relation for the isothermal compressibility, $\kappa_{T}^p$, at $p$ is
\begin{equation}
    \kappa_{T}^{p} = \frac{\kappa_{T}^{p_0}}{v_0[1-\frac{2a}{v_0^2} \kappa_{T}^{p_0}]},
\end{equation}

where $\kappa_{T}^{p_0}$ is the isothermal compressibility at $p_0$ such that
\begin{equation}
\begin{split}
    \kappa_{T}^{p_0} = - \frac{\partial \ln v_0}{\partial p} = -\frac{1}{v_0} \frac{\partial v_0}{\partial p} = -\frac{1}{v_{0}}\ \sum_j \frac{\partial f_j v_{j,0}}{\partial p} \\
    = -\frac{1}{v_{0}}\ \sum_j \left[f_j \frac{\partial v_{j,0}}{\partial p} +v_{j,0}\frac{\partial f_j}{\partial p} \right] \equiv \sum_j \kappa_{T,j}^{p_0}.
\end{split}
\end{equation}
Here $\kappa_{T,j}^{p_0}$ is the isothermal compressibility of $j^{th}$ state at $p_0$. 

Similarly, the isobaric expansion coefficient, $\alpha_p$, at $p$ is 
\begin{equation}
    \alpha_{p} = \frac{\alpha_{p_0}}{v_0[1-\frac{2a}{v_0^2} \kappa_{T}^{p_0}]},
\end{equation}

where $\alpha_{p_0}$ is given by
\begin{equation}
\begin{split}
    \alpha_{p_0} =  \frac{\partial \ln v_0}{\partial T} = \frac{1}{v_{0}}\ \sum_j \left[f_j \frac{\partial v_{j,0}}{\partial T} +v_{j,0}\frac{\partial f_j}{\partial T} \right] \equiv \sum_j \alpha_{p_0,j}.
\end{split}
\end{equation}
Here $\alpha_{p_0,j}$ is the expansion coefficient of $j^{th}$ state at $p_0$. 

Finally the heat capacity, $C_p$ at $p$ is given by
\begin{equation}\label{cp_hp}
    C_p = \frac{\partial h}{\partial T}\Bigr|_p = \frac{\partial h}{\partial T} \Bigr|_{p_0} +\frac{\partial h}{\partial p_0}\Bigr|_T \frac{\partial p_0}{\partial T} \Bigr|_p
\end{equation}

It can also be shown that turning on attraction reduces the enthalpy from $h$ at pressure $p$ to $h_{p_0} - 2a/v_0$ at pressure $p_0$ \cite{Truskett2002a}. So, it can be shown that Eq. \ref{cp_hp} leads to 
\begin{equation}
    C_p = C_{p_0} + \frac{2a \alpha_{p_0} \kappa_T^{p_0} p_0}{v_0 [1-\frac{2a}{v_0^2} \kappa_{T}^{p_0}]}.
\end{equation}
Where $C_{p_0}$ is 
\begin{equation}
    C_{p_0} =\sum_j \frac{\partial  (f_j \langle h_{p_0,j} \rangle)}{\partial T} = \sum_j \left(f_j \frac{\partial \langle h_{p_0,j} \rangle}{\partial T} +\langle h_{p_0,j} \rangle \frac{\partial f_j }{\partial T} \right) \equiv \sum_j C_{p_0,j},
\end{equation}
here $h_{p_0,j}$ and $C_{p_0,j}$ are enthalpy and heat capacity of $j^{th}$ state at $p_0$, respectively.

\newpage

\subsection*{Model parameters}
To find the optimum parameters we used PyGAD \cite{Gad2021}, a genetic algorithm Python library, to search the multidimensional bound parameter space to minimize the error between predicted and selected experimental data \cite{Mishima2010, Urbic2018, Mallamace2014, Angell1982, Chen1977, Lin2012, Pallares2016}. Table \ref{tb:paratable} lists the optimized parameters used in this study. All parameters are rounded off to 3 significant digits. \\
\begin{table}[!ht]
\caption{Parameters of the current model}
\vspace{0.5cm}
\begin{center}
\begin{tabular}{ |c|c|c| }
 \hline
 \textbf{Parameters} & \textbf{Description} & \textbf{Values} \\[0.1ex]
 \hline
 $r_{Ih}$           & Equilibrium HB length in Ih ice         & 2.77 \AA \\[0.5ex]
 $r_{Cage}^{max}$   & Cooperative HB length maximum           & 2.97 \AA \\[0.5ex]
 $r_{pHB}^{max}$    & Pair HB length maximum                  & 3.23 \AA \\[0.5ex]
 $r_{vdW}^{max}$    & van der Waals bond length maximum       & 3.53 \AA \\[0.5ex]
 $r_{d}$            & Inaccessible hardcore size of water      & 2.63 \AA \\[0.5ex]
 $x_{pHB}$          & pHB water volume perturbation factor    & 0.724 \\[0.5ex]
 $x_{vdW}$          & vdW water volume perturbation factor    & 1.29 \\[0.5ex]
 $x_{NI}$           & Water vapor volume perturbation factor  & 1700 \\[0.5ex]
 $\epsilon_{Cage}^0$  & Cooperative HB maximum interaction potential & 20.5 kJmol$^{-1}$ \\[0.5ex]
 $\epsilon_{pHB}^0$   & Pair HB maximum interaction potential & 19.2 kJmol$^{-1}$ \\[0.5ex]
 $\epsilon_{vdW}^0$   & Pair vdW maximum interaction potential        & 10.3 kJmol$^{-1}$ \\[0.5ex]

 $\kappa_{x}$       & Translational spring constant     & 13.1 kJmol$^{-1}$ {\AA}$^{-2}$ \\[0.5ex]
 $\kappa_{\theta}$  & Rotational spring constant      & 897 kJmol$^{-1}$ \\[0.5ex]
 $a$                & Dispersion interaction constant between cages & 274 kJ{\AA}$^3$mol$^-1$ \\[0.5ex]
 \hline
\end{tabular}\label{tb:paratable}
\end{center}
\end{table}

All the water-water bond length parameters (2.63 \AA - 3.53 \AA) and hydrogen bond parameters are within the experimental range. This model has 14 adjustable parameters which are similar in number to the total of 12 parameters (see Table \ref{tb:tip&opc}) of the four-point models like TIP4P/05 and OPC.\\
\begin{table}[!ht]
\caption{Parameters of the fixed charge atomistic models TIP4P/05 and OPC}
\begin{center}
\begin{tabular}{ |c|c|c|c| }
 \hline
 \textbf{Parameters} & \textbf{Description} & \textbf{TIP4P/05}\cite{Abascal2005} & \textbf{OPC}\cite{Izadi2014}   \\[0.5ex]
 \hline
 $\epsilon_O$   & LJ dispersion energy of O  & 0.63627 kJ/mol & 0.89036 kJ/mol \\[0.5ex]
 $\epsilon_H$   & LJ dispersion energy of H  & 0.0 kJ/mol & 0.0 kJ/mol \\[0.5ex]
 $\epsilon_M$   & LJ dispersion energy of M  & 0.0 kJ/mol & 0.0 kJ/mol \\[0.5ex]
 $\sigma_{O}$ & LJ size of O     & 3.1589 \AA     & 3.16655 \AA     \\[0.5ex]
 $\sigma_{H}$ & LJ size of H     & 0.0 \AA        & 0.0 \AA     \\[0.5ex]
 $\sigma_{M}$ & LJ size of M     & 0.0 \AA        & 0.0 \AA     \\[0.5ex]
 $q_O$        & Charge on O      & 0.0 $e$        & 0.0 $e$          \\[0.5ex]
 $q_M$        & Charge on M      & -1.1128 $e$    & -1.3582 $e$   \\[0.5ex]
 $q_H$        & Charge on H      & 0.5564 $e$     & 0.6791 $e$    \\ [0.5ex]
$r_OH$       & OH distance           & 0.9572 \AA     & 0.8724 \AA     \\[0.5ex]
 $r_{OM}$     & OM distance           & 0.1546 \AA     & 0.1594 \AA     \\[0.5ex]
 $\angle HOH$ & HOH angle             & 104.52$^\circ$ & 103.6$^\circ$ \\ [0.5ex]
 \hline
\end{tabular}\label{tb:tip&opc}
\end{center}
\end{table}

\clearpage
\section*{Phase transition and critical point in the supercooled water}

Recent experiments indicate the existence of liquid-liquid phase transition (LLPT) \cite{Kim2020, pathak2021}; however, huge fluctuation in pressure (from ambient to 350 MPa) makes the location of the liquid-liquid critical point still elusive (LLCP).  Figure 7 (in the main article) shows that the current model predicts the LLCP at 176 MPa and 159 K, where extrema in $C_p^{max}, \kappa_T^{max}$, and $\alpha_p^{min}$ collapse on the WL. It also shows the corresponding LLPT and the binodal line. This predicted LLCP is within the range of previously estimated or predicted LLCP using molecular simulations and thermodynamic analyses \cite{Poole1992, Mishima1998, Poole2005, Liu2009, Cuthbertson2011, Yamada2002, Paschek2005, Paschek2008, Abascal2010, Corradini2010, Palmer2014, Debenedetti2020, Gartner2022b, Weldon2024, Sciortino2025, dhabal2024, Coronas2024a, Hestand2018, Ni2016, Mishima2023}. The range of predicted critical pressure, $P_c \approx (60-360)$ MPa, and critical temperature, $T_c \approx (172-250)$ K, is vast. Mishima and Sumita \cite{Mishima2023} recently predicted $P_c \approx 105 \pm 9$ MPa by extrapolating empirically fitted $n^{th}$ order polynomials ($n = 3..10$) to calculate the pressure from volume and temperature data, $p_{cal} = \sum_{i=0}^n c_iV^iT^i$. However, their prediction is contingent upon the order of polynomials and experimental data used for fitting. Additionally, the possibility of over-fitting can't be overruled with 64 adjustable parameters (for the representative $7 \times 7$ polynomial). Whereas recent analyses of atomistic simulation data \cite{Kim2017, Spah2018, Hestand2018} that predict $P_c$ in the vicinity of $\approx 100$ MPa, compare $\kappa_T^{max}$ prediction with experimental value at ambient pressure. They assume that the predicted $\kappa_T^{max}$ value would be the same as the experimental $\kappa_T^{max}$ at the same pressure distance from their respective critical points. Often, these models perform poorly at capturing water's volumetric and thermodynamic properties beyond ambient pressure. For example, a sharper $\kappa_T^{max}$ peak leads to a lower critical pressure compared to a broader peak \cite{Kim2017, Spah2018, Caupin2019}. Predictions based on analyses oblivious of such effects of water's properties on the location of LLCP could be misleading. 

Recent $ab\ initio$ derived water model, rWAIL \cite{Weldon2024}, and DNN surrogate of MB-pol (DNN@MB-pol \cite{Sciortino2025}) place the $P_c$ in the vicinity of 100 MPa. In addition to liquid water configurations, these models are optimized to predict highly tetrahedral ice polymorph structures \cite{Weldon2024, Sciortino2025}. Angell and Kapko \cite{Angell2016} showed that stronger tetrahedrality leads to a lower critical pressure, which could explain a significantly lower prediction of $P_c$ by models calibrated to reproduce ice structures. Further, Hestand and Skinner \cite{Hestand2018} noted that several parameters, including but not limited to charges \cite{Lascaris2016}, bond flexibility \cite{Smallenburg2015} and tetrahedrality \cite{Angell2016}, and nuclear quantum effect \cite{Nguyen2018}, could influence the location of LLCP of water and similar tetrahedral liquid. This suggests that the predictions of LLCP by water models could be inherently biased towards the choice of experimental properties prioritized in the validation and training schemes.

In comparison, the current model is purely validated against the macroscopic liquid water's experimental data. It predicts water's response function in excellent quantitative agreement with experiments across temperature and pressure (see figures 2 - 5 of the main article). Liquid water is significantly less tetrahedrally ordered than different ice polymorphs. Hence, a slightly higher critical pressure ($P_c = 176$ MPa) prediction by the current model compared to DNN@MB-pol and rWAIL. Extremely slow speed and small system size of such $ab\ initio$ based models make the validation of other response functions ($C_p$, $\kappa_T$, and $\alpha_p$) beyond the ambient pressure against experiments extremely difficult. Whereas, the extremely fast nature of our model allows us to quickly recalibrate and help pinpoint the location of LLCP as more experiments become available in the erstwhile \emph{no man's land} in the future.

\clearpage

\section*{Supporting Figures}
\begin{figure}[!htb]
\centerline{\includegraphics[width=3.0in]{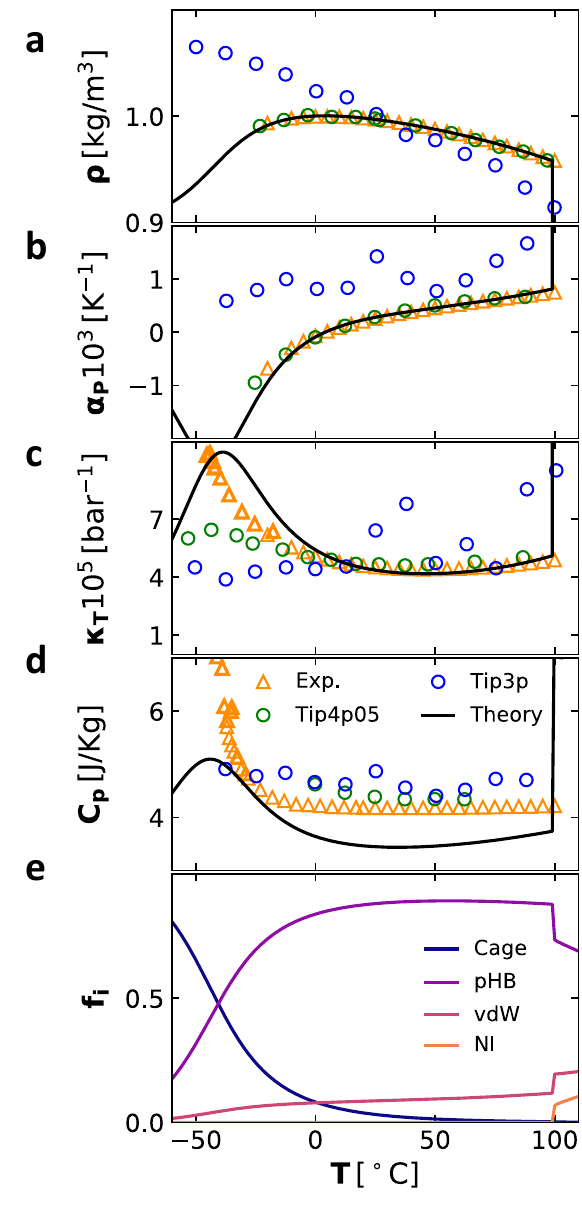}}
\caption[]{\label{fig:vle} \textbf{Comparison with experiments and explicit water model predictions.}  Thermophysical properties and population showing first order liquid-vapor transition at 1 atm and 100$^\circ$C. Theory predictions are shown with lines, experiments with open orange triangles, TIP3P with open blue circles, and TIP4P05 with open green circles. Current predictions are much better than TIP3P water, which is still commonly used in biophysical simulations.}
\end{figure}

\begin{figure}[!htb]
\centerline{\includegraphics[width=3.0in]{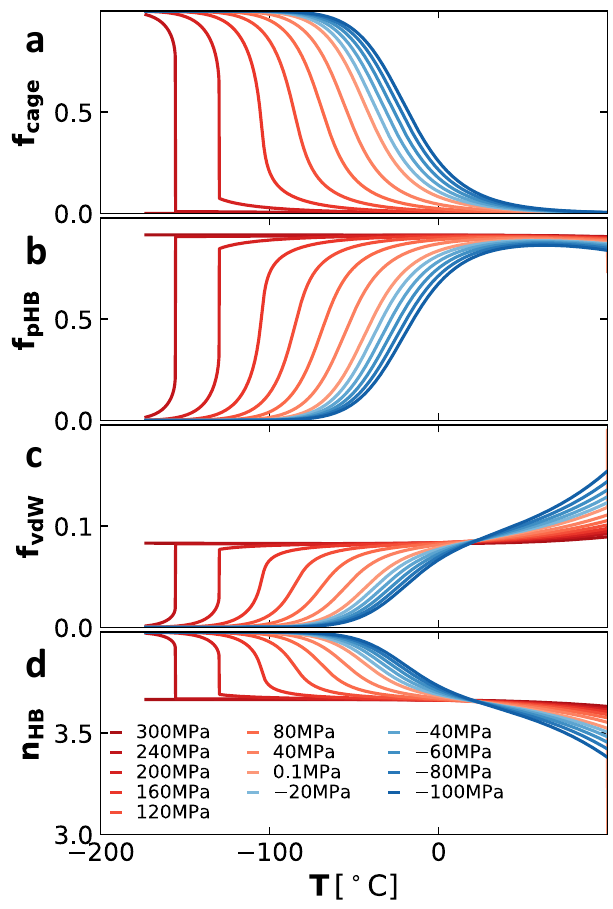}}
\caption[]{\label{fig:populations} \textbf{Isobars of microstates' populations and average hydrogen bond versus temperature.} \textbf{a)} cage state. \textbf{b)} pHB state. \textbf{c)} vdW state. \textbf{d)} average hydrogen bonds. $p>0$ isobars are in shades of red. $p<0$ isobars are in shades of blue. The Isobar of three states at constant pressure are shown with the same color in both \textbf{a)} and \textbf{b)}.}
\end{figure}

\begin{figure}[!htb]
\centerline{\includegraphics[width=4.0in]{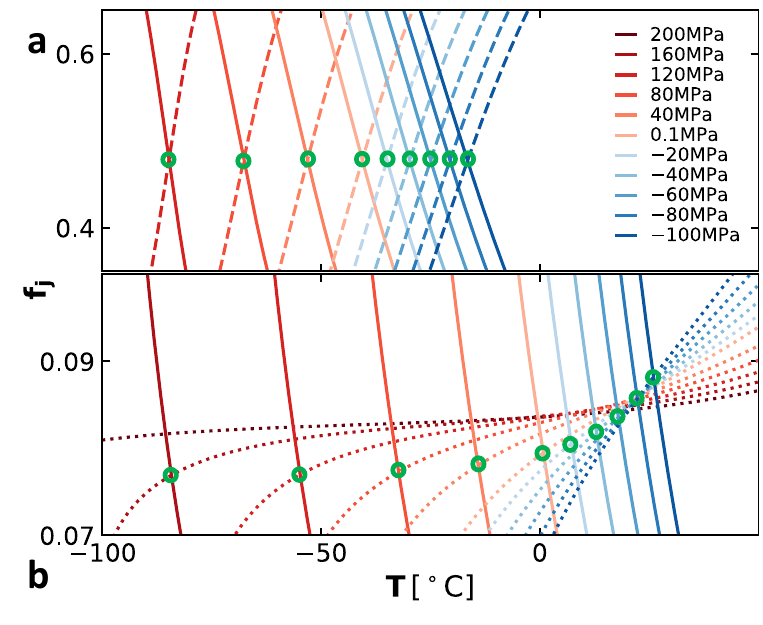}}
\caption[]{\label{fig:cross} \textbf{Crossovers between microstates.} \textbf{a)} Crossovers between $f_{cage}$ (solid lines) and $f_{pHB}$ (dashed lines). \textbf{b)} crossovers between $f_{cage}$ (solid lines) and $f_{vdW}$ (dotted line). The crossover points are shown with green open circles.}
\end{figure}

\begin{figure}[!htb]
\centerline{\includegraphics[width=6.0in]{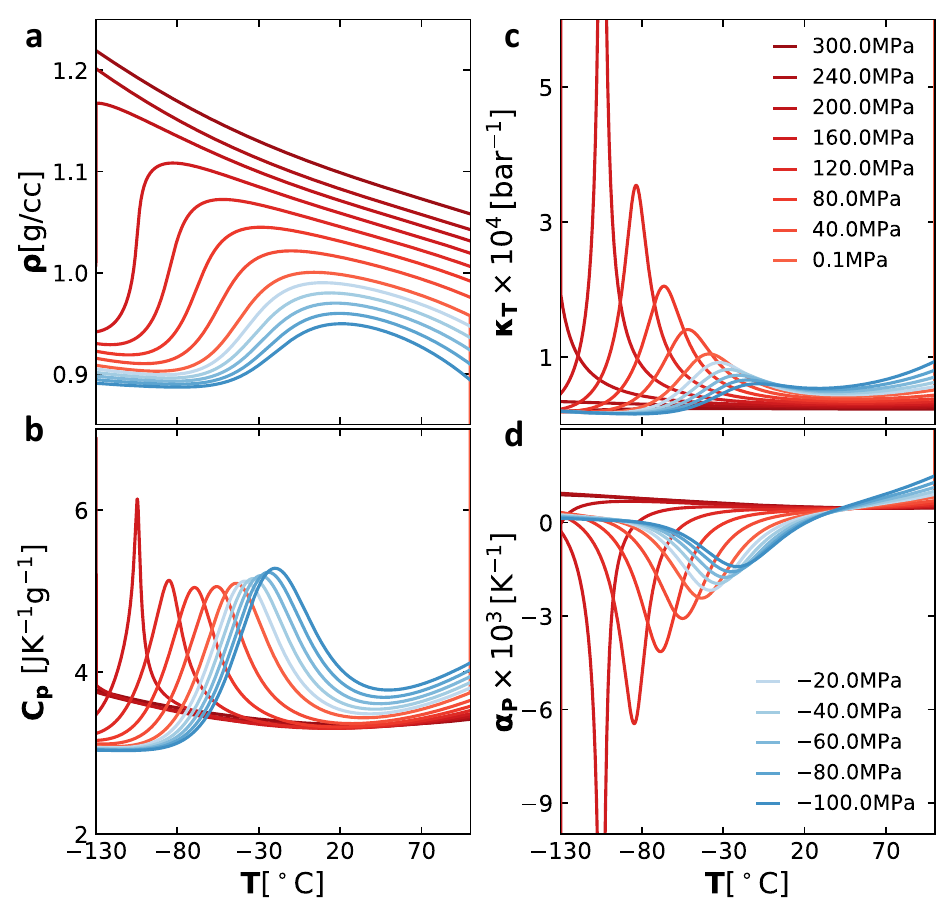}}
\caption[]{\label{fig:isobars} \textbf{Isobars predictions of volumetric and energetic responses.} \textbf{a)} Density, $\rho$, \textbf{b)} Heat capacity,$C_p$. \textbf{c)} Isothermal compressibility, $\kappa_T$, and \textbf{d)} Expansion coefficient, $\alpha_p$. $p>0$ isobars are in shades of red. $p<0$ isobars are in shades of blue. The maximum density, $\rho^{max}$, and peaks in response functions $C_p^{max}$, $\kappa_T^{max}$, and $\alpha_p^{min}$ shift to a lower temperature with pressure in the supercooled region.}
\end{figure}

\begin{figure}[!htb]
\centerline{\includegraphics[width=3.2in]{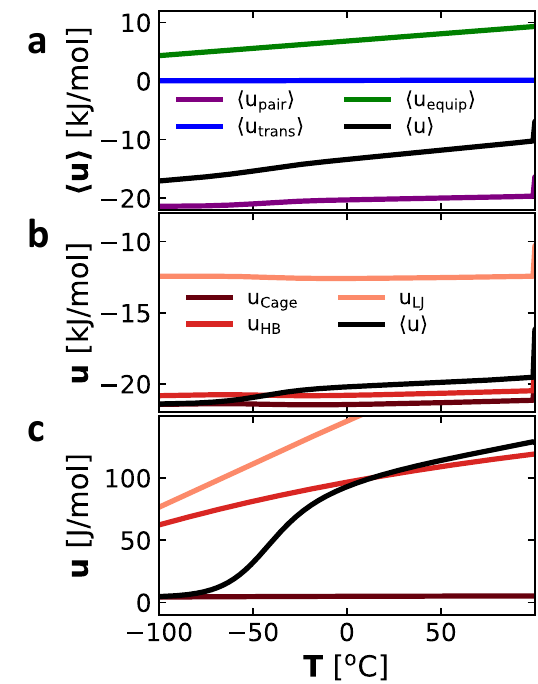}}
\caption[]{\label{fig: bond_ene} \textbf{Average temperature dependent energy contributions per water-water bond.} \textbf{a)} Ensemble averaged contributions from pair interaction $\langle \epsilon \rangle$ in purple, translation $\langle u_{trans} \rangle$ in blue, equipartition $\langle u_{equip} \rangle$ in green, and total $\langle u \rangle$ in black. \textbf{(b, c)} Energy contribution from different water states; ensemble average $\langle u \rangle$ is in black and state contributions are shown with shades of red. \textbf{b)} Without thermal contribution, $6k_BT$ equipartition to bonds. \textbf{c)} Transnational contributions.}
\end{figure}

\clearpage
\newpage
\bibliography{Supplementary.bbl}